\documentclass[aps,prl,reprint,amsmath,amssymb,superscriptaddress]{revtex4-1}

\usepackage{graphicx}
\usepackage{bm}
\usepackage{dcolumn}
\usepackage{gensymb}
\usepackage[utf8]{inputenc}
\usepackage{float}

\usepackage{url,hyperref,breakurl}

\usepackage{array,mathtools,amssymb,booktabs}
\newcolumntype{C}{>{$}c<{$}}
\AtBeginDocument{
\heavyrulewidth=.08em
\lightrulewidth=.05em
\cmidrulewidth=.03em
\belowrulesep=.65ex
\belowbottomsep=0pt
\aboverulesep=.4ex
\abovetopsep=0pt
\cmidrulesep=\doublerulesep
\cmidrulekern=.5em
\defaultaddspace=.5em
}

\usepackage{hyperref}
\newcommand{\tref}[1]{Table~\ref{#1}}
\newcommand{\eref}[1]{Eq.~(\ref{#1})}

\begin{document}

\title{Precision calculation of hyperfine constants for extracting nuclear moments of $^{229}$Th}

\author{S. G. Porsev}
\affiliation{Department of Physics and Astronomy, University of Delaware, Newark, Delaware 19716, USA}
\affiliation{Petersburg Nuclear Physics Institute of NRC ``Kurchatov Institute'', Gatchina, Leningrad District 188300, Russia}
\author{M. S. Safronova}
\affiliation{Department of Physics and Astronomy, University of Delaware, Newark, Delaware 19716, USA}
\author{M. G. Kozlov}
\affiliation{Petersburg Nuclear Physics Institute of NRC ``Kurchatov Institute'', Gatchina, Leningrad District 188300, Russia}
\affiliation{St. Petersburg Electrotechnical University LETI, St. Petersburg 197376, Russia}

\begin{abstract}
Determination of nuclear moments for many nuclei relies on the computation of hyperfine constants, with theoretical uncertainties directly affecting the resulting uncertainties of the nuclear moments. In this work we improve the precision of such method by including for the first time an iterative solution of equations for the core triple cluster amplitudes into the relativistic coupled-cluster method, with large-scale complete basis sets.
We carried out calculations of the energies and magnetic dipole and electric quadrupole hyperfine structure constants for the low-lying states
of $^{229}$Th$^{3+}$ in the framework of such relativistic coupled-cluster single double triple (CCSDT) method. We present a detailed study of
various corrections to all calculated properties. Using the theory results and experimental data we found the nuclear magnetic dipole and electric quadrupole moments to be $\mu = 0.366(6) \mu_N$ and $Q = 3.11(2)\, e{\rm b}$, and reducing the uncertainty of the quadrupole moment by a factor of three. The Bohr-Weisskopf effect of the finite nuclear magnetization is investigated, with bounds placed on the  deviation of the magnetization distribution from the uniform one.
\end{abstract}

\date{\today}

\maketitle

% ====================
%\section{Introduction}
% ====================
Laser spectroscopy in combination  with atomic structure calculations can be used to directly determine
nuclear moments, in a nuclear theory independent way. Such an approach is limited by the ability to calculate hyperfine structure (HFS)
constants $A$ and $B$ from first principles to a high precision. This problem is exacerbated in heavy atoms where electronic correlation corrections increase.
%together with relative increase of the higher-order contributions.
A separate problem is a determination of the accuracy of the theoretical computations, as the theory uncertainty directly contributes to the uncertainty of the extracted nuclear magnetic-dipole and electric-quadupole moments.
While numerical uncertainties can be generally determined,
%the uncertainties due to ``missing physics'' are difficult to estimate and require  good understanding of relative importance of various contributions %and method accuracy.
an estimate of other uncertainties is difficult because it requires a good understanding of relative importance of various contributions
and method accuracy. In addition, theoretical computations require modeling of magnetization distribution, which is generally not known.

In this work, we consider a solution of these problems for the $^{229}$Th nucleus, motivated by its unique features described below.
The development of precision methods for extracting nuclear moments from laser spectroscopy measurements becomes of particular  importance now as more rare isotopes will become available with high yield at the Facility for Rare Isotope Beams (FRIB) \cite{Abel19} for exploring nuclear physics
properties, especially of actinides.

As it was established by more than 40 years ago, the nuclear transition frequency
between the ground and first excited state of $^{229}$Th is unusually small and amounts to only several eV~\cite{KroRei76}.
Subsequent measurements of this quantity confirmed it; the current most precise value of $8.19(12)$ eV ~\cite{PeiSchSaf21} is an average of two recent measurements \cite{SeiWenBil19,SikGeiHen20}.
 Such a unique feature of this isotope opens up a number of theoretical and experimental research opportunities.
Special interest in this nuclear transition is motivated by a possibility to build a superprecise nuclear clock~\cite{PeiTam03}
and very high sensitivity to the effects of possible temporal variation of the fundamental constants, including the fine structure
constant $\alpha$, strong interaction, and quark mass~\cite{Fla06,FadBerFla20}.

The present uncertainty in the nuclear transition frequency 0.12 eV, corresponding to $\sim$30 THz, is many orders of magnitude greater than the natural linewidth, expected to be in the mHz range. To determine the nuclear transition frequency with laser spectroscopic precision, as well as other properties of the ground and isomeric nuclear states, further experimental and theoretical investigations are required~\cite{PeiSchSaf21}.
%The nuclear magnetic and quadrupole moments, which are the key properties for the nuclear states, were determined in Ref.~\cite{SafSafRad13}.
Using the experimentally measured and theoretically calculated HFS constants $A$ and $B$, the
nuclear magnetic dipole and electric quadrupole moments were determined in Ref.~\cite{SafSafRad13} to be $\mu = 0.360(7) \mu_N$
and $Q = 3.11(6)\, e{\rm b}$ (where $\mu_N$ is the nuclear magneton and $e$ is the elementary charge). This value of the magnetic moment contradicts
to the result $\mu = 0.46(4) \mu_N$ found in Ref.\cite{GerLueVer74} and to the recent nuclear calculation value 0.43-0.48 $\mu_N$ obtained
in Ref.~\cite{MinPal21}.
Motivated by a necessity to confirm the results of Ref.~\cite{SafSafRad13} and by a need to better understand the Th$^{3+}$ properties
for the development of the nuclear clock~\cite{PeiSchSaf21}, we further developed the coupled-cluster single double triple (CCSDT) method,
fully including into consideration both valence and core triple excitations and applied it for the high-accurate calculation of
the $^{229}$Th$^{3+}$ properties. The same approach can be used for other systems. A combination of the CCSDT method with configuration
interaction (see Ref.~\cite{SafKozJoh09})
%(see \cite{SafKozJoh09} for description of the linearized LCCSD+CI method)
can also be applied  to allow more accurate extraction of nuclear moments for systems with more then one valence electron.

\begin{figure}[t]
  \includegraphics[scale=0.4]{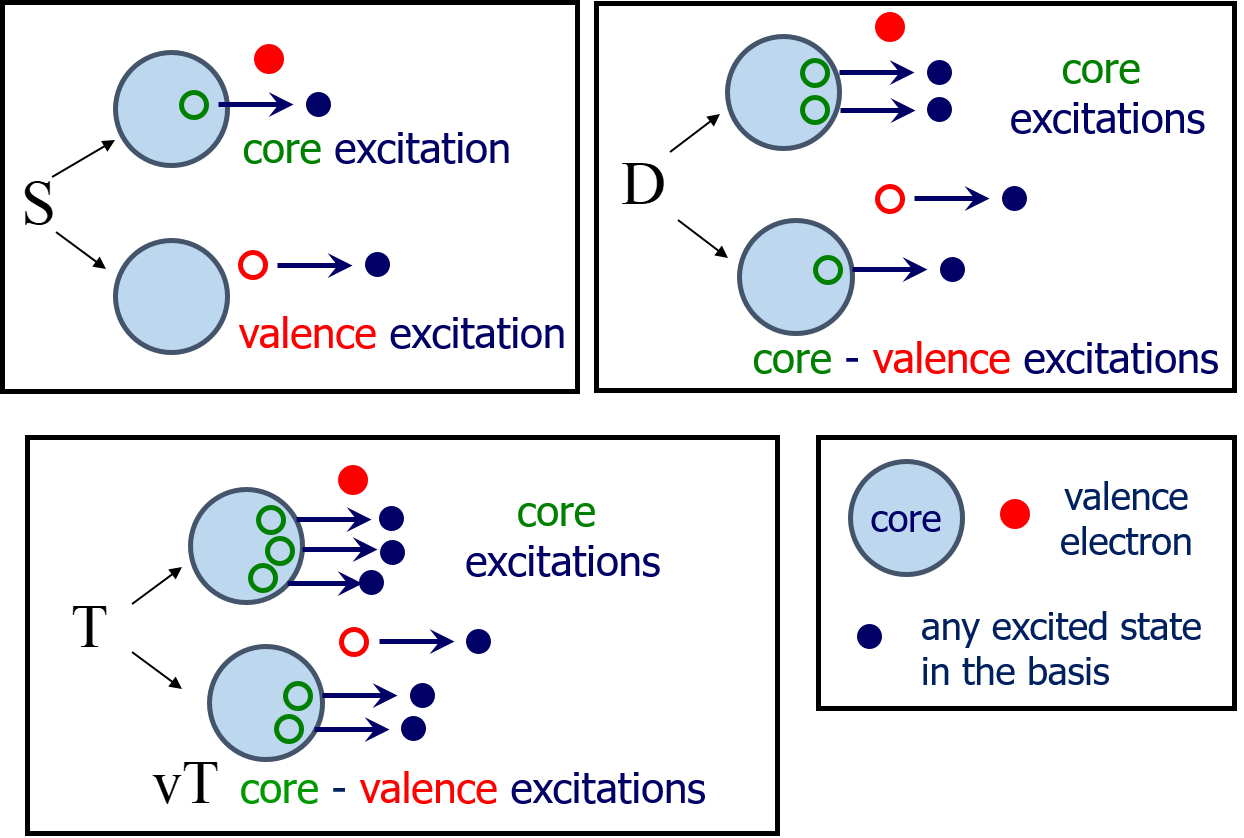}
  \caption{Illustration of single (S), double (D), and triple (T) excitations in the coupled-cluster approach.}
 \label{fig1}
 \end{figure}

The simplest version of this approach, the linearized coupled-cluster single double (LCCSD) method,
was developed in Ref.~\cite{BluJohSap91}. In this version only the linear terms involving the single ($S$) and double ($D$) excitations
of the valence and core electrons are included into consideration. A wide range of properties of univalent systems can be
calculated with a very good accuracy using the LCCSD method (see, e.g.,~\cite{SafJoh08}). But systematic
prediction of the properties of transition matrix elements and HFS constants with an accuracy below 1\% requires an inclusion of the terms
beyond LCCSD, i.e., the nonlinear (NL) terms and the triple excitations.
The quadratic NL single double terms are combinations of the single and double excitations of the electrons
that take the form of $S^2$, $SD$, $D^2$ combinations.
Triple valence and core terms involve excitations of two core electrons and a valence electron and three core electrons, respectively.
Schematically single, double, and triple excitations are presented in Fig.~\ref{fig1}.

The NL terms and/or valence triple (vT) excitations were included into consideration in the leading order in a number of
works~\cite{EliKalIsh96,RupSafJoh07,SafSafRad13,ArnChaKae19,Sah06,SahDasCha07}.
A more sophisticated approach, where the equations for the {\it valence} triples were solved iteratively, was developed and applied
in~\cite{PorDer06,PorBelDer09,PorBelDer10}. A detailed study of contributions of the NL terms and valence triples
in calculating the energies, magnetic dipole HFS constants and $E1$ transition amplitudes in Ba$^+$
was presented in the recent work~\cite{PorSaf21}, which stressed the need for inclusion of the core triple terms.

Aiming to improve the calculation accuracy of the HFS constants and, respectively, to reduce uncertainty of $\mu$ and $Q$, we made the next
step in developing the method of calculation and included the core triple excitations into consideration. The equations for the core triples
were solved {\it iteratively}. A very detailed discussion devoted to including the vT excitations into consideration
in the frame of the coupled cluster method, was presented in Ref.~\cite{PorDer06}. The inclusion of the core triples can be done in the same manner.
%Technically, in the equations for the valence triples (see Eqs.~(18-20) in Ref.~\cite{PorDer06}) one needs to change the index of the valence
%electron to the index of the core electron. Similar change can be done in other terms including the valence triples.
To the best of our knowledge, due to exceptional complexity of the problem and very high computational
demands, the core triples were never included into computation of the properties of complicated atomic systems.
%We also accelerated the computation of valence triples by a factor of 10 allowing inclusion of larger number of valence triple excitations
%associated with higher partial waves.
%We note that this problem is so challenging for a heavy system with a large core, that we were unable to fully implement it even using modern computer facilities. As we discuss in more detail below, not all excitations of the core electrons were included in calculation of the core triple amplitudes.

%We carried out calculations of the energies of the lowest-lying states and their HFS constants $A$ and $B$ in the framework of the
%coupled-cluster single double triple (CCSDT) method. In addition to
%the LCCSD and NL terms, the valence and core triple excitation coefficients were included by solving the respective equations iteratively.
%Combining our results with the experimental values for the constants $A$ and $B$~\cite{SafSafRad13}, we found the nuclear magnetic dipole
%and electric quadrupole moments to be $\mu = 0.366(6) \mu_N$ and $Q = 3.11(2)\, e{\rm b}$, confirming the results of Ref.~\cite{SafSafRad13}
%and reducing the uncertainty of the quadrupole moment by three times.

% =============================
{\bf Method of calculation.}
%\label{method}
% =============================
We evaluated the energies and HFS constants $A$ and $B$ of the lowest-lying states using a version of the high-precision relativistic coupled-cluster method developed in Ref.~\cite{PorDer06} and augmented by inclusion of the core triple excitations.
% To the best of our knowledge, it was for the first time, when a full-scale calculation of properties of a heavy ion included the iterative
% solution of the equations for the core triples.
Quadratic non-linear (NL) terms ($S^2$, $SD$, and $D^2$) were also included in the equations for singles and doubles, but omitted in the equations
for triples. The cubic and higher-order terms were also omitted as these are expected to be small.

We consider Th$^{3+}$ as a univalent ion.
%and construct the basis set in the $V^{N-1}$ approximation, where $N$ is the number of electrons.
The initial Dirac-Hartree-Fock (DHF) self-consistency procedure included the Breit interaction and was done for the core electrons.
%Then the $5f_{5/2,7/2}$ and $6d_{3/2,5/2}$
%orbitals were constructed in the frozen-core potential. The remaining virtual orbitals were
%formed using 40 basis set B-spline orbitals of order 7 defined on a nonlinear grid with 500 points.
%This basis set included partial waves with the orbital quantum number up to $l= 6$.
The coupled cluster equations were solved in a basis set consisting of single-particle states.
In the equations for singles, doubles,
and valence triples the sums over excited states were carried out with 35 basis orbitals with orbital quantum number $l \leq 6$.
Due to high computational demands to the iterative solution of the equations for the core triple excitations, certain restrictions were applied.
%We solved these equations allowing the core electrons excitations from the $[4d-6p]$ core shells, the maximal orbital
%quantum number
%, $l^{\rm tr}_{\rm max}$,
%of all excited orbitals was equal to 5, and the largest principal quantum number
%$n^{\rm tr}_{\rm max}$
%of the virtual orbitals where excitations were allowed was 25.
%The equations for the core triples were solved under the same conditions but
%the maximal orbital quantum number of all excited orbitals was equal to 4.
But, as our estimates showed, the not-included electrons excitations
%(to the orbitals with $l=6$ and with $n > 25$ for any $l$)
can contribute to the removal energies of the valence states only at the level of several tens of cm$^{-1}$ and
can change the HFS constants at the level of 0.1\%.
Thus, we included the triple excitations into consideration practically in full.

% -------------------
{\bf Energies.}
% -------------------
Numerical results for the energies are presented in Table~\ref{Tab:E}. The lowest-order DHF contribution to the energies
(with the inclusion of the Breit interaction) is labeled ``BDHF''. At the next step, we carried out calculation in the linearized coupled-cluster
single double (LCCSD) approximation. Then, we subsequently added the NL terms, valence and core triples into consideration, designating these
calculations as CCSD, CCSDvT, and CCSDT, respectively. Thus, each subsequent calculation includes all terms taken into account at the previous
stage and the additional terms specific for the present approximation. In this way, the most complete calculation is carried out
in the CCSDT approximation.

The removal energies of the valence states obtained on each stage are presented in the table.
We also found complementary correction due to the basis extrapolation ($\Delta E_{\rm extrap}$), determined as  the contribution of
the higher ($l > 6$) partial waves. Based on an empiric rule obtained for Ag-like ions in Ref.~\cite{SafDzuFla14PRA1} and used in
Ref.~\cite{PorSaf21}, we estimate this contribution
as the difference of two calculations carried out with $l_{\rm max} = 6$ and $l_{\rm max} = 5$.
For the $5f$ and $6d$ states the quantum-electrodynamic corrections to the energies are small and we did not
include them into the full-scale calculation. As our estimate shows this effect can potentially change the removal energies of
the considered states at the level of 100-200 cm$^{-1}$.
The total values, presented in the row labeled ``$E_{\rm total}$'', are found as $E_{\rm CCSDT} + \Delta E_{\rm extrap}$.

The experimental removal energy for the ground state is
$231065\,(200)$ cm$^{-1}$~\cite{RalKraRea11}, i.e., its uncertainty is comparable with the difference between our total value and the
experimental result. The experimental values for the excited states were taken from Ref.~\cite{ThIV}.

To illustrate a consistent improvement in the results when we successively add different coupled-cluster terms,
we present the differences between the theoretical and experimental values obtained at each stage in the low panel of Table~\ref{Tab:E}.
Comparing $\Delta_{\rm CCSDT}$ and $\Delta_{\rm LCCSD}$, we see that the difference between the theory and experiment decreased by almost
four times for the $5f$ states and by two times for the $6d$ states when we included the NL terms and triples into consideration.
% ####################################################################
\begin{table}[h]
\caption{The removal energies of the $5f_{5/2;7/2}$ and $6d_{3/2;5/2}$ states for Th$^{3+}$ (in cm$^{-1}$) in
different approximations, discussed in the text, are presented. The theoretical total and experimental results
are given in the rows $E_{\rm total}$ and $E_{\rm expt}$.
The difference between the total and experimental values is presented (in \%) in the row labeled ``Diff. (\%)''.}
\label{Tab:E}
\begin{ruledtabular}
\begin{tabular}{lrrrr}
\smallskip
                                        & $5f_{5/2}$ & $5f_{7/2}$ &  $6d_{3/2}$    & $6d_{5/2}$ \\
\hline \\[-0.7pc]
$E_{\rm BDHF}$                          &  207310    &   203393   &    211842      &   207686   \\[0.2pc]
$E_{\rm LCCSD}$                         &  232308    &   227978   &    222871      &   217543   \\[0.2pc]
$E_{\rm CCSDT}$                         &  231640    &   227307   &    222490      &   217174   \\[0.2pc]
$E_{\rm CCSDvT}$                        &  230819    &   226538   &    222472      &   217259   \\[0.2pc]
$E_{\rm CCSDT}$                         &  230693    &   226398   &    222268      &   217032   \\[0.2pc]
$\Delta E_{\rm extrap}$                 &    1055    &     1032   &      257       &      242   \\[0.4pc]

$E_{\rm total}$                         &  231748    &   227431   &    222526      &   217274   \\[0.3pc]
$E_{\rm expt}$~\cite{RalKraRea11,ThIV}  &  231065    &   226740   &    221872      &   216579   \\[0.1pc]
Diff. (\%)                              &    0.30    &     0.30   &      0.29      &     0.32   \\
\hline \\[-0.6pc]
${\Delta_{\rm LCCSD}}^{\rm a}$          &    2298    &     2271   &      1256      &     1205   \\[0.2pc]
${\Delta_{\rm CCSD}}^{\rm b}$           &    1630    &     1599   &       875      &      836   \\[0.2pc]
${\Delta_{\rm CCSDvT}}^{\rm c}$         &     809    &      831   &       857      &      922   \\[0.2pc]
${\Delta_{\rm CCSDT}}^{\rm d}$          &     683    &      691   &       654      &      695
\end{tabular}
\end{ruledtabular}
\begin{flushleft}
$^{\rm a}\!\Delta_{\rm LCCSD} \equiv E_{\rm LCCSD} + \Delta E_{\rm extrap} - E_{\rm expt}$; \\
$^{\rm b}\!\Delta_{\rm CCSD} \equiv E_{\rm CCSD} + \Delta E_{\rm extrap} - E_{\rm expt}$; \\
$^{\rm c}\!\Delta_{\rm CCSDvT} \equiv E_{\rm CCSDvT} + \Delta E_{\rm extrap} - E_{\rm expt}$; \\
$^{\rm d}\!\Delta_{\rm CCSDT} \equiv E_{\rm CCSDT} + \Delta E_{\rm extrap} - E_{\rm expt}$.
\end{flushleft}
\end{table}
%####################################################################
For completeness, using the total values of the removal energies, we present in Table~\ref{Tab:E1} the theoretical transition energies
counted from the ground state and compare them with the experimental data~\cite{ThIV}.
% ####################################################################
\begin{table}[h]
\caption{The theoretical and experimental~\cite{ThIV} transition energies (in cm$^{-1}$) of the excited states counted from the ground state.
The differences between the experiment and theory are given in columns 4 and 5 in cm$^{-1}$ and \%.}
\label{Tab:E1}
\begin{ruledtabular}
\begin{tabular}{lcccc}
\smallskip
                                    &  Theory    & Experiment & Diff. & Diff. \%\\
\hline \\[-0.7pc]
$5f_{5/2}$                          &     0      &     0      &       &        \\[0.2pc]
$5f_{7/2}$                          &   4318     &   4325     & 7     &  0.16  \\[0.2pc]
$6d_{3/2}$                          &   9223     &   9193     & -30   & -0.33  \\[0.2pc]
$6d_{5/2}$                          &  14475     &  14486     & 11    &  0.08
\end{tabular}
\end{ruledtabular}
%\begin{flushleft}
%\end{flushleft}
\end{table}
%####################################################################

% ----------------------------------------
%\subsection{HFS constants}
{\bf Hyperfine structure constants.}
% ----------------------------------------
The magnetic dipole and electric quadrupole HFS constants $A$ and $B$ were calculated for the low-lying states of $^{229}$Th$^{3+}$ in
Refs.~\cite{SafSafRad13,BerDzuFla09}.
%Based on these calculations and the experimental data for the constants~\cite{CamRadKuz11}, the
%nuclear magnetic dipole and electric quadrupole moments were found to be $\mu = 0.360(7)\,\mu_N$ and $Q= 3.11(6)\,e{\rm b}$~\cite{SafSafRad13}.
In Ref.~\cite{SafSafRad13} the authors also used for the calculation the coupled-cluster method, but a less sophisticated version.
In particular, the NL terms and core triples were disregarded while the valence triples were
included perturbatively.
%In certain cases the final results were found by scaling the {\it ab initio} results.
%The scaling factors were determined by using the reference benchmark systems for which the nuclear moments were known very precisely.
In this work we carry out the more complete calculation including the NL terms and the valence and core triple excitations.
In addition, our calculation is pure {\it ab initio}; no semi-empirical methods are applied.

The results for the magnetic-dipole HFS constants $A_t \equiv A/g$ (where $g = (\mu/\mu_N)/I$ is the $g$ factor and $I$ is the nuclear spin, $I=5/2$)
are presented in \tref{Tab:Ahfs}.
%###############################################################################################################################################
\begin{table}[h]
\caption{Different contributions to $A_t$ (in MHz) and obtaining the recommended value of $g$ are
explained in the text. The experimental values of the HFS constants $A$~\cite{CamRadKuz11} are given in the row
labeled ``$A$(experim.)''.
%The values of $\mu_I$ are obtained as the ratios of $A$(experim.) and values listed in the row labeled ``Total''.
%The recommended value of $\mu_I$ is given in the row labeled ``$\mu_I ({\rm recommended})$''.
The uncertainties are given in parentheses.}
\label{Tab:Ahfs}
\begin{ruledtabular}
\begin{tabular}{lcccc}
                                 & $5f_{5/2}$   & $5f_{7/2}$  & $6d_{3/2}$    & $6d_{5/2}$ \\
\hline \\[-0.6pc]
BDHF                             &   507        &   263       &   831         &    304         \\[0.4pc]

$\Delta$(SD)                     &    72        &   -45       &   268         &   -386         \\[0.1pc]
LCCSD                            &   579        &   218       &  1099         &    -81         \\[0.4pc]

$\Delta$(NL)                     &    -3.3      &    -4.6     &   -17         &     18         \\[0.1pc]
$\Delta$(vT)                     &   -12        &    -5.1     &   -21         &    -46         \\[0.1pc]
$\Delta$(cT)                     &    -1.5      &    -2.3     &     5.8       &    -1.8        \\[0.1pc]
CCSDT                            &   562        &   206       &  1067         &   -111         \\[0.3pc]

Basis extrap.                    &    -0.2      &    2.6      &    -4.5       &      7.1       \\[0.3pc]

Total                            &  $562(3)$    &  $209(3)$   &  $1063(12)$   &    -$104(22)$    \\[0.2pc]
Ref.~\cite{SafSafRad13}$^{\rm a}$&   573        &   215       &   1079        &     -92        \\[0.4pc]
\hline \\[-0.6pc]
$A$(experim.)~\cite{CamRadKuz11} &   $82.2(6)$  &  $31.4(7)$  &  $155.3(1.2)$ &   -$12.6(7)$   \\[0.4pc]
$g ({\rm recommended})$          &  \multicolumn{4}{c}{$0.1465(24)$}                            \\[0.2pc]
Ref.~\cite{SafSafRad13}          &  \multicolumn{4}{c}{$0.1440(28)$}
%
% BDHF                             &   203        &   105       &   332         &    122         \\[0.4pc]
%
% $\Delta$(SD)                     &    29        &   -18       &   107         &   -154         \\[0.1pc]
% LCCSD                            &   232        &    87.3     &   440         &    -33         \\[0.4pc]
%
% $\Delta$(NL)                     &    -1.3      &    -1.8     &    -6.8       &     7.1        \\[0.1pc]
% $\Delta$(vT)                     &    -4.9      &    -2.0      &   -8.2       &   -18          \\[0.1pc]
% $\Delta$(cT)                     &    -0.6      &    -0.9     &     2.3       &    -0.7        \\[0.1pc]
% CCSDT                            &   225        &    82.5     &   427         &    -45         \\[0.3pc]
%
% Basis extrap.                    &    -0.1      &    1.0      &    -1.8       &      2.8       \\[0.3pc]
%
% Total                            &  $225(1)$    & $83.5(1.1)$ &  $425(5)$     &    -$42(9)$    \\[0.2pc]
% Ref.~\cite{SafSafRad13}          &   229.2      &   86.1      &   431.5       &    -36.7       \\[0.4pc]
% \hline \\[-0.6pc]
% $A$(experim.)~\cite{CamRadKuz11} &   $82.2(6)$  &  $31.4(7)$  &  $155.3(1.2)$ &   -$12.6(7)$   \\[0.4pc]
% %$\mu_I$                          &  $0.366(3)$  & $0.376(10)$ &  $0.365(5)$   &    $0.302$     \\[0.3pc]
% $\mu_I ({\rm recommended})$      &  \multicolumn{4}{c}{$0.366(6)$}                             \\[0.2pc]
% Ref.~\cite{SafSafRad13}          &  \multicolumn{4}{c}{$0.360(7)$}
\end{tabular}
\end{ruledtabular}
\begin{flushleft}
$^{\rm a}$The values, listed in Ref.~\cite{SafSafRad13}, are multiplied by $I=5/2$.
\end{flushleft}
\end{table}
%#############################################################################################################################################

The LCCSD and BDHF values and the difference between them,  $\Delta$(SD), are given in the upper panel of the table.
The rows 4-6 give the corrections due to the NL terms, $\Delta$(NL), and the valence and core triples, $\Delta$(vT) and $\Delta$(cT).
The CCSDT values, obtained as the sum of the LCCSD values and the NL, vT, and cT corrections, are presented in the row
labeled ``CCSDT''. As seen, $\Delta$(cT) is several times less in absolute value than $\Delta$(vT) for all considered states.
The basis extrapolation corrections are given in the row labeled ``Basis extrap.''.
%accounts for the contribution of the partial waves with $l > 6$ and
%the contribution of the orbitals with principal quantum number $n > 40$.
The total values, listed in the row ``Total'', are found as the sum of the CCSDT value and the basis extrapolation correction.

Based on a comparison of the theoretical and experimental HFS constants for a number of univalent elements, the authors of
Ref.~\cite{SafSafRad13} suggested a method to estimate the uncertainties of these constants. The uncertainty of the $A$ and $B$ calculations,
carried out in the framework of the couple-cluster method, is expected to be on the order of 3-6\% of the total correlation correction
(found as the difference between the final and LCCSD value), if this correction does not exceed 50\%.
Following this approach, we estimate the uncertainties of $A_t$ for the $5f_{5/2}$, $5f_{7/2}$, and $6d_{3/2}$ states
as 5\% of their total correlation corrections. For the $6d_{5/2}$ state the correlation correction is very large and
the validity of this method is questionable. Applying it, we roughly estimate the uncertainty of $A_t$ for this state
at the level of 20-25\%.

We note that all values presented in \tref{Tab:Ahfs} were obtained for the nucleus considered as the charged ball with uniform
magnetization. But, as it was shown in Ref.~\cite{MinPal17}, the nucleus of $^{229}$Th has a complex structure and different collective
effects, such as quadruple-octupole vibration-rotation motion of the nucleus, the single-particle motion of the
unpaired nucleon and the Coriolis interaction between this nucleon and the nuclear core, are important.
Thus, a real nuclear magnetization can differ from the uniform magnetization. To investigate this problem,
we follow the approach developed in Ref.~\cite{DemKonIma21}. We can express the HFS constant $A$ as,
\begin{eqnarray}
  A = g\, A_0\, (1 - d_{\rm nuc}\, y) ,
\label{A1}
\end{eqnarray}
where $A_0$ is the theoretical value calculated at the point-like magnetization of the nucleus and
$d_{\rm nuc}$ and $y$ are the parameters depending on the nuclear and electronic structure, respectively.
The quantities $g$ and $d_{\rm nuc}$ are assumed to be unknown, the experimental value of $A$ can be used on the left hand side
of \eref{A1}, and $y$ can be found from the calculation.

Indeed, taking into account that $d_{\rm nuc}=0$ and $d_{\rm nuc}=1$
correspond to the point-like and uniform magnetization, respectively, we can find $y$ from \eref{A1} as
\begin{eqnarray}
  y = 1 - A_t/A_0,
\label{A_T}
\end{eqnarray}
where $A_t$ are given in \tref{Tab:Ahfs} for different states.
We note that the ratios $A_t/A_0$ are very stable and insensitive to different corrections and we determine
the uncertainty of $y$ at the level of 0.02\%.

To find $g$ and $d_{\rm nuc}$, we use the HFS constants for the $5f_{5/2}$ and $6d_{3/2}$ states, which are known most precisely both
experimentally and theoretically. Using for each of them Eq.~(\ref{A1}) and solving the system of two equations
in two unknowns, we arrive at
\begin{eqnarray}
\label{g_d}
 g &\approx& 0.1465(24),
% \nonumber
\\
 d_{\rm nuc} &\approx& 1.7(2.1) .
\label{dnuc}
\end{eqnarray}

Thus, the $g$ factor is determined with the accuracy $\sim 1.5\%$. Using this value, we find the nuclear magnetic moment,
$\mu_I = g I \approx 0.366(6)$. This result is in good agreement with that reported in Ref.~\cite{SafSafRad13}, $\mu_I \approx 0.360(7)$.
We note that the uncertainty estimate in \cite{SafSafRad13} did not include uncertainty due to the magnetization distribution.

%Though the calculation accuracy of $d_{\rm nuc}$ is poor, we conclude that it is not very large. According to our calculation, we can place the following limits
%$$-0.4 \leq d_{\rm nuc} \leq 3.8.$$
For heavy nuclei the parameter $d_\mathrm{nuc}$ can vary over a wide range. For example, for the gold isotopes $^{197,193,191}$Au
with the nuclear spin of 3/2, $d_\mathrm{nuc}=-5.5(6)$~\cite{DemKonIma21}. Our result \eqref{dnuc} suggests that for ${}^{229}$Th the absolute value of $d_\mathrm{nuc}$ is smaller and it is most likely positive. As a result, the correction to the $g$ factor, due to inhomogeneity of the nuclear
magnetization, is small but not negligible. For determining the magnetic dipole nuclear moment $\mu_I$ with the accuracy better than 1\%, this
effect should be taken into account.

% Using the experimental results for the HFS constants~\cite{CamRadKuz11} and theoretical total values of $A/\mu_I$,
% we are able to find the values of $\mu_I$. They obtained as the ratios of $A$(experim.) and values listed in the row labeled
% ``Total'' in \tref{Tab:Ahfs}. We note that all four values, found in such a way, agree to each other within their uncertainties.
% The nuclear magnetic moment $\mu_I$, obtained from the experimental and theoretical results for the $5f_{5/2}$ state, has the least
% uncertainty and we consider it as the recommended one.

In \tref{Tab:Bhfs} we present the results obtained for the electric-quadrupole HFS constants $B/Q$. All designations in the upper panel of the
table are the same as in \tref{Tab:Ahfs}. In the lower panel, we present the experimental results for the HFS constants $B$~\cite{CamRadKuz11}.
The values of $Q$ (in $e$b) are found as the ratios of $B$(experim.) and values listed in the row labeled ``Total''.
The recommended value of $Q$ (in $e$b) is given in the row labeled ``$Q$ (recommended)''.
%###############################################################################################################################################
\begin{table}[h]
\caption{Different contributions to the electric-quadrupole HFS constants $B/Q$ (in MHz/($e$b)) for $^{229}$Th$^{3+}$,
explained in the text, are presented. The experimental values of the HFS constants $B$~\cite{CamRadKuz11} are given in the row
labeled ``$B$(experim.)''. The values of $Q$ (in $e$b) are obtained as the ratios of $B$(experim.) and values listed in the row labeled
``Total''; the recommended value of $Q$ (in $e$b) is given in the row labeled ``$Q$ (recommended)''.
The uncertainties are given in parentheses.}
\label{Tab:Bhfs}
\begin{ruledtabular}
\begin{tabular}{lcccc}
                                    & $5f_{5/2}$   & $5f_{7/2}$  & $6d_{3/2}$    & $6d_{5/2}$ \\
\hline \\[-0.6pc]
BDHF                                &   535        &   572       &   611         &    648         \\[0.4pc]

$\Delta$(SD)                        &   202        &   251       &   132         &    228         \\[0.1pc]
LCCSD                               &   737        &   822       &   743         &    877         \\[0.4pc]

$\Delta$(NL)                        &    38        &    45       &     9         &      9         \\[0.1pc]
$\Delta$(vT)                        &   -55        &   -57       &   -34         &    -27         \\[0.1pc]
$\Delta$(cT)                        &     3        &     3       &     7         &      7         \\[0.1pc]
CCSDT                               &   723        &   814       &   725         &    866         \\[0.3pc]

Basis extrap.                       &     6        &     8       &     3         &      4         \\[0.3pc]

Total                               & $729(10)$    &  $822(13)$  &  $728(6)$     &  $869(11)$    \\[0.2pc]
Ref.~\cite{SafSafRad13}             &  725         &   809       &   738         &   873         \\[0.2pc]
Ref.~\cite{BerDzuFla09}             &  740         &   860       &   690         &   860         \\
\hline \\[-0.6pc]
$B$(experim.)~\cite{CamRadKuz11}    & $2269(2)$    & $2550(12)$  &  $2265(9)$    &  $2694(7)$   \\[0.4pc]
$Q$                                 & $3.11(4)$    & $3.10(5)$   &  $3.11(3)$    &  $3.10(4)$   \\[0.3pc]
$Q ({\rm recommended})$             &  \multicolumn{4}{c}{$3.11(2)$}                            \\[0.2pc]
Ref.~\cite{SafSafRad13}             &  \multicolumn{4}{c}{$3.11(6)$}                            \\[0.2pc]
Refs.~\cite{CamRadKuz11,BerDzuFla09}&  \multicolumn{4}{c}{$3.11(16)^{\rm a}$}
\end{tabular}
\end{ruledtabular}
\begin{flushleft}
$^{\rm a}${This result was obtained using the measurements of Ref.~\cite{CamRadKuz11} and calculations of Ref.~\cite{BerDzuFla09}.}
\end{flushleft}
\end{table}
%#############################################################################################################################################

The values of $B/Q$ obtained in this work turned out to be between the results of Ref.~\cite{SafSafRad13} and Ref.~\cite{BerDzuFla09}
but are somewhat closer to the former. The experimental uncertainty for the constants $B$ does not exceed 0.5\% while the theoretical uncertainties
are at the level of 0.8-1.5\%. The uncertainties were determined in the same manner as it was done for the HFS constants $A$.
The constants $B$ are large for all four considered states and their fractional uncertainties are
comparable. For this reason, the recommended value was obtained as the weighted average over four values of $Q$ given in \tref{Tab:Bhfs}.
We note the perfect agreement of our recommended value with the results obtained in Refs.~\cite{SafSafRad13,CamRadKuz11,BerDzuFla09}, but our
uncertainty is a few times smaller.

% ====================================
%\section{Conclusion and final remarks}
{\bf Conclusion.}
% ====================================
We carried out the pure {\it ab initio} calculations of the energies and HFS constants $A$ and $B$ for the low-lying states
in the framework of the relativistic CCSDT method. We have determined the different contributions to these quantities,
including the contributions from the quadratic NL terms and the valence and core triples, and the basis set extrapolation correction.
The equations for the core triples were solved {\it iteratively}. To the best of our knowledge, it was done for the first time in
calculating the properties of a heavy element.

Using the theoretical values obtained in this work and the experimental results for the HFS constants~\cite{SafSafRad13},
we determined the values of the nuclear magnetic dipole and electric quadrupole moments.
Analyzing the results obtained at the different stages we determined the uncertainties of the recommended values to be 1.5\% for $\mu_I$
and 0.6\% for $Q$. We investigated the effect of the inhomogeneity of the nuclear magnetization, and found it to be small but not negligible.
We conclude that it should be taken into account to determine the magnetic dipole nuclear moment $\mu_I$ with the accuracy better than 1\%.

Further improvement of the theoretical results is possible by applying
new parallel atomic code allowing to use hundreds of processors simultaneously. It will give us a possibility to include the core triples
into consideration more consistently. Experimental efforts to measure the HFS (especially, magnetic dipole) constants more precisely, are also
needed to improve precision further and better understand the effect of nuclear magnetization.

{\bf Acknowledgments.}
We are grateful to Yu. Demidov for useful discussion.
This work is a part of the ``Thorium Nuclear Clock'' project that  has received funding from the European Research Council  (ERC) under
the European Union's Horizon 2020 research and innovation program (Grant Agreement No. 856415).
S.P. and M.K. acknowledge support by the Russian Science Foundation under Grant No. 19-12-00157.

%\bibliography{./ThIV}

%merlin.mbs apsrev4-1.bst 2010-07-25 4.21a (PWD, AO, DPC) hacked
%Control: key (0)
%Control: author (72) initials jnrlst
%Control: editor formatted (1) identically to author
%Control: production of article title (-1) disabled
%Control: page (0) single
%Control: year (1) truncated
%Control: production of eprint (0) enabled
%

\end{document}